# Comparative study of under-expressed prognostic biomarkers and pivotal signaling pathways in colon cancer and ulcerative colitis using integrated bioinformatics approach.

Dr. Sedigheh Behrouzifar

Department of Medical Sciences, Shahrood Branch,

Islamic Azad University, Shahrood, Iran

sedighehbehrouzifar@gmail.com

## Abstract

*Colon cancer is a prevalent gastrointestinal malignancy arising in the colon. Ulcerative colitis(UC) is one of the risk factors of colorectal cancer. The detection of under-expressed biomarkers and molecular mechanisms in UC and colon cancer can lead to effective management of colitis-associated cancer.*

*A total of two mRNA expression datasets (GSE87473 and GSE44076) were downloaded from the Gene Expression Omnibus (GEO) database. GSE87473 contains 21 healthy samples, 27 extensive ulcerative colitis samples and 60 limited ulcerative colitis samples. GSE44076 contains 98 colon cancer samples and 98 healthy samples. GEO2R was used to screen differentially expressed genes (DEGs) between extensive ulcerative colitis samples and healthy samples, limited ulcerative colitis samples and healthy samples, and colon cancer samples and healthy samples.*

*The inclusion criteria for DEGs included an adjusted p-value <0.05 and a log(2) fold change $\leq$-2. Venn diagram of DEGs was depicted for every three groups (extensive ulcerative colitis, limited ulcerative colitis and colon cancer). Protein-protein interaction (PPI) network of DEGs in every three groups was constructed using STRING online database. The DEGs of each group were imported to Cytoscape software separately and the hub genes were screened. Then, the Enrichr web server was used to perform KEGG (Kyoto Encyclopedia of Genes and Genomes) enrichment analyses and adjusted p-value<0.05 was considered statistically significant. Furthermore, Kaplan-Meier curve in GEPIA (http://gepia.cancer-pku.cn/) was used for analyzing the overall survival (OS) of hub genes.*

*In extensive ulcerative colitis, limited ulcerative colitis and colon cancer groups, 95,69 and 635 under-expressed genes with adjusted p-value<0.05 and log(2) fold change$\leq$-2 were detected respectively. Using Cytoscape software, the genes with degree$\geq$ 15 including CLCA1, SLC26A3, SI, KIT, HPGDS, NR1H4, ADIPOQ, PPARGC1A, GCG, MS4A12, GUCA2A and FABP1 were screened as hub under-expressed genes in colon cancer. In extensive ulcerative colitis, the genes with degree$\geq$5 including ABCB1, ABCG2, UGT1A6, CYP2B6 and AQP8 were identified as hub genes. Moreover, the genes including NR1H4, CYP2B6, ABCB1, ABCG2, UGT2A3 and PLA2G12B were detected as hub genes with degree$\geq$5 in limited ulcerative colitis. According to inclusion criteria and venn diagram, the downregulated gene NR1H4 was common gene in limited ulcerative colitis and colon cancer. The genes MS4A12 and GUCA2A were common genes in extensive ulcerative colitis and colon cancer. The genes CLCA1, SLC26A3, SI, KIT, HPGDS, ADIPOQ, PPARGC1A, PPARGC1A, GCG and FABP1 were*

*distinctive genes in colon cancer. The genes UGT1A6 and CYP2B6 were common genes in extensive ulcerative colitis and limited ulcerative colitis, and the genes ABCB1, ABCG2, AQP8, and UGT2A3 were common genes in three groups. According to KEGG enrichment analysis, hub under-expressed genes in colon cancer were enriched in the pathways including Pancreatic secretion, Mineral absorption, Drug metabolism, Metabolism of xenobiotics by cytochrome P450, Arachidonic acid metabolism, Bile secretion, PPAR signaling pathway, Adipocytokine signaling pathway and cAMP signaling pathway. The hub genes in extensive and limited ulcerative colitis were enriched in Bile secretion, ABC transporters, Steroid hormone biosynthesis, Retinol metabolism, Metabolism of xenobiotics by cytochrome P450, Drug metabolism, Chemical carcinogenesis and Fat digestion and absorption. According to survival analysis, CLCA1 (Calcium-activated chloride channel regulator 1), PPARGC1A (Peroxisome proliferator-activated receptor gamma coactivator 1-alph) and AQP8 (Aquaporin-8) were with poor overall survival. The current in silico study showed that downregulation of CLCA1, PPARGC1A and AQP8 genes may increase cancer cell invasion and metastasis ability. The recent researches showed that CLCA1 overexpression inhibited colorectal cancer aggressiveness, and overexpression of AQP8 reduced cell proliferation, migration and invasion in colon cancer. The role of downregulation of PPARGC1A gene in poor survival of patients with colon cancer has not been revealed yet.*

**1- Introduction:**

Ulcerative colitis (UC), is a chronic inflammatory disorder of the large intestine. UC was first defined in the 1800s by Samuel Wilks. Prevalence of UC is increasing in Asia (1). When the diagnosis is made, most UC patients are between the ages of 30 and 40. The clinical features of the disease include abdominal pain, diarrhea, and hematochezia (2). Approximately 50% of UC patients don't have an advancing disease course (3). Patients with UC have a 2.4-fold higher risk of colon cancer than the general population (4). Colorectal cancer (CRC) is the third most prevalent cancer worldwide (5).

The main risk factors for colon cancer are high fat diet and low fiber diet, and genetic factors may be involved in the progression of colon cancer (6). In spite of decades of effort, the intricate pathogenesis of UC is not yet fully identified (7). According to recent studies, the application of non-steroidal anti-inflammatory drugs may decrease the probability of UC transforming into colorectal cancer (8). Moreover, recent advances in biologically targeted therapies have resulted in improved disease management, and surgery is needed in only some patients (9). The treat-to-target approach has shifted the goal of UC treatment from symptomatic management to support targeting therapy to prevent long-term disease complications (10). Also, remarkable progress has been made in the development of biomarker-driven therapies for patients with multiple cancer types. However, precision oncology for patients with colorectal cancer is controversial (11). Biomarkers play an important role in assessing and management of diseases. Hence, identifying predictor biomarkers of disabling diseases is important to beginning a target-based treatment.

Exploring the mechanisms underlying the progression of colon cancer and achieving medications or drug combinations that can stop advancement of UC toward colon cancer would be required. Therefore, identifying and screening key target proteins (biomarkers) as well as insight into the biological pathways involved in the pathogenesis and progression of disease in order to improve response to treatment is important.

## 2- Materials and methods:

A total of two mRNA expression datasets (GSE87473 and GSE44076) were downloaded from the Gene Expression Omnibus (GEO) database. GSE87473 contains 21 healthy samples, 27 extensive ulcerative colitis samples and 60 limited ulcerative colitis samples. GSE44076 contains 98 colon cancer samples and 98 healthy samples. GEO2R was used to screen differentially expressed genes (DEGs) between extensive ulcerative colitis samples and healthy samples, limited ulcerative colitis samples and healthy samples, and colon cancer samples and healthy samples.

The inclusion criteria for DEGs included an adjusted p-value <0.05 and a log(2) fold change $\leq$-2. Venn diagram of DEGs was depicted for every three groups (extensive ulcerative colitis, limited ulcerative colitis and colon cancer). Protein-protein interaction (PPI) network of DEGs in every three groups was constructed using STRING online database. The DEGs of each group were imported to Cytoscape software separately and the hub genes were screened. Then, the Enrichr web server was used to perform KEGG (Kyoto Encyclopedia of Genes and Genomes) enrichment analyses and adjusted p-value<0.05 was considered statistically significant. Furthermore, Kaplan-Meier curve in GEPIA (http://gepia.cancer-pku.cn/) was used for analyzing the overall survival (OS) of hub genes.

## 3- Results:

In extensive ulcerative colitis, limited ulcerative colitis and colon cancer groups, 95,69 and 635 under-expressed genes with adjusted p-value<0.05 and log(2) fold change$\leq$-2 were detected respectively. Using Cytoscape software, the genes with degree$\geq$ 15 including CLCA1, SLC26A3, SI, KIT, HPGDS, NR1H4, ADIPOQ, PPARGC1A, GCG, MS4A12, GUCA2A and FABP1 were screened as hub under-expressed genes in colon cancer (Figure 1). In extensive ulcerative colitis, the genes with degree$\geq$5 including ABCB1, ABCG2, UGT1A6, CYP2B6 and AQP8 were identified as hub genes (Figure 2). Moreover, the genes including NR1H4, CYP2B6, ABCB1, ABCG2, UGT2A3 and PLA2G12B were detected as hub genes with degree$\geq$5 in limited ulcerative colitis (Figure 3).

According to inclusion criteria and venn diagram (Figure 4), the downregulated gene NR1H4 was common gene in limited ulcerative colitis and colon cancer. The genes MS4A12 and GUCA2A were common genes in extensive ulcerative colitis and colon cancer. The genes CLCA1, SLC26A3, SI, KIT, HPGDS, ADIPOQ, PPARGC1A, PPARGC1A, GCG and FABP1 were distinctive genes in colon cancer. The genes UGT1A6 and CYP2B6 were common genes in extensive ulcerative colitis and limited ulcerative colitis, and the genes ABCB1, ABCG2, AQP8 and UGT2A3 were common genes in three groups. According to KEGG enrichment analysis, hub under-expressed genes in colon cancer were enriched in the pathways including Pancreatic secretion, Mineral absorption, Drug metabolism, Metabolism of xenobiotics by cytochrome P450, Arachidonic acid metabolism, Bile secretion, PPAR signaling pathway, Adipocytokine signaling pathway and cAMP signaling pathway. The hub genes in extensive and limited ulcerative colitis were enriched in Bile secretion, ABC transporters, Steroid hormone biosynthesis, Retinol metabolism, Metabolism of xenobiotics by cytochrome P450, Drug metabolism, Chemical carcinogenesis and Fat digestion and absorption. According to survival analysis, CLCA1, PPARGC1A and AQP8 were with poor overall survival (Figure 5).

**4- Discussion:**

In the present study, Calcium-activated chloride channel regulator 1 (CLCA1), Chloride anion exchanger (SLC26A3), Sucrase-isomaltase, intestinal (SI), Mast/stem cell growth factor receptor Kit (KIT), Hematopoietic prostaglandin D synthase (HPGDS), Bile acid receptor (NR1H4), Adiponectin (ADIPOQ), Peroxisome proliferator-activated receptor gamma coactivator 1-alpha (PPARGC1A), Pro-glucagon (GCG), Membrane-spanning 4-domains subfamily A member 12 (MS4A12), Guanylin (GUCA2A) and Fatty acid-binding protein (FABP1) exhibited the highest interaction degree in colon cancer. The recent researches showed that CLCA1 overexpression inhibited colorectal cancer aggressiveness (12). In agreement with the present study, Beata and colleagues identified CLCA1, SLC26A3 and FABP1 as hub genes that were downregulated in colorectal carcinoma (13). One recent research revealed that CLCA1 acts as a tumor suppressor gene by suppressing Bcl-2 overexpression. In this study, there was a correlation between downregulation of CLCA1 and poorer prognosis in colorectal carcinoma (14). In the study of Bo Yang and colleagues, knockdown of CLCA1 with siRNA considerably inhibited cell differentiation and promoted cell proliferation in Caco-2 cultures (15). In present study, KEGG enrichment analysis showed that CLCA1 gene was significantly enriched in the pathway "pancreatic secretion".

The results of one study uncovered that, SLC26A3 overexpression can inhibit growth of cancer cell lines in vitro (16). According to KEGG enrichment analysis, SLC26A3 gene was enriched in the pathways including "pancreatic secretion" and "mineral absorption".

Yamazaki and colleagues reported a correlation between poorer prognosis and lower expression levels of FABP1 in liver metastasis (17). Moreover, in the study of Kheirelseid and colleagues, the downregulation of FABP1 in colorectal carcinogenesis has been reported (18). According to KEGG enrichment analysis, FABP1 gene was enriched in "PPAR signaling pathway".

Koichi Iwanaga and colleagues exhibited that under-expression of HPGDS aggravated colitis and accelerated tumor formation in mice (19). According to KEGG enrichment analysis, HPGDS gene was enriched in the pathways including "drug metabolism", "metabolism of xenobiotics by cytochrome P450" and "arachidonic acid metabolism".

In the study of Yun Jeong Lee and colleagues, NR1H4 KO cells showed impaired cell proliferation and increased apoptotic cell death compared to control colon cancer cells (20). Based on the present study, this gene was common gene in limited ulcerative colitis and colon cancer. NR1H4 gene was enriched in the pathway "bile secretion".

A low serum level of ADIPOQ is associated with an increased risk of various types of malignancies including colorectal cancer (21). According to recent studies, ADIPOQ is inversely associated with tumor grade in colorectal cancer patients (22, 23, 24). In the present study, ADIPOQ gene was enriched in the pathways including "PPAR signaling pathway" and "adipocytokine signaling pathway".

In a recent study, GCG was found to be significantly downregulated in colon and rectum adenocarcinomas compared with healthy control samples. GCG gene was shown to be related to the prognostic outcomes of this cancer (25). Based on the present study, GCG gene was enriched in the "cAMP signaling pathway".

MS4A12 participates in cell membrane composition, cell differentiation, proliferation, and cell cycle regulation (26). low expression of MS4A12 had a poor survival, suggesting that MS4A12 might be a molecular marker for diagnosis and prognosis (27). One recent study exhibited that PPARGC1A acts as a tumor suppressor in hepatocellular carcinoma (28). Based on the present study, PPARGC1A gene was enriched in the pathway "adipocytokine signaling pathway".

Recent studies have demonstrated a link between GUCY2C silencing and intestinal dysfunction and tumorigenesis. GUCA2A binds and activates GUCY2C to regulate proliferation and metabolism in intestine (29). In the present study, the genes MS4A12 and GUCA2A were common genes in extensive ulcerative colitis and colon cancer. In the recent research of Jing Han and colleagues, primary colorectal cancer patients with lower expression of CLCA4 and MS4A12 had poorer overall survivals (26).

According to the study of Hui Zhang and colleagues, GUCA2A can be a candidate marker of poor prognosis in patients with colorectal cancer and further studies are needed to detect molecular mechanism through which GUCA2A plays a role in colorectal cancer (30).

The present study revealed that the genes UGT1A6 and CYP2B6 were common genes in extensive ulcerative colitis and limited ulcerative colitis, and the genes ABCB1, ABCG2, AQP8 and UGT2A3 were common genes in three groups. In contrast to present study, a recent study showed that UGT2A3 gene was as top ten upregulated genes in ulcerative colitis (31).

Based on the present study, UGT1A6 and UGT2A3 genes were enriched in the pathways including "bile secretion", "steroid hormone biosynthesis", "retinol metabolism", "metabolism of xenobiotics by cytochrome P450", "ascorbate and aldarate metabolism", "pentose and glucuronate interconversions", "drug metabolism" and "porphyrin and chlorophyll metabolism". The present study showed that CYP2B6 gene was enriched in the pathways including "Retinol metabolism", "Metabolism of xenobiotics by cytochrome P450" and "Drug metabolism". Furthermore, ABCB1 (MDR1) and ABCG2 genes were enriched in the pathways including "bile secretion" and "ABC transporters". Recently, single nucleotide polymorphisms (SNPs) of multi-drug resistance 1 (MDR1) gene were found to be related to the pathogenesis of ulcerative colitis (32). A recent study exhibited that AQP8 expression is reduced in the ileum and colon of UC patients (33). In another recent study, overexpression of AQP8 reduced colorectal cancer cell proliferation, migration and invasion in vitro (34). Based on the present study, AQP8 gene was enriched in the pathway "bile secretion".

In the present study, overall survival of key genes in colon cancer, extensive and limited ulcerative colitis was evaluated by Kaplan-Meier curve. CLCA1, PPARGC1A and AQP8 were with poor overall survival. The role of downregulation of PPARGC1A gene in poor survival of patients with colon cancer has not been revealed yet.

**Conclusion:**

In present study, novel prognostic genes and key biological pathways were uncovered in colon cancer, extensive and limited ulcerative colitis using systems biology approach. Hence, performing in vitro and in vivo researches to verify the results is essential.

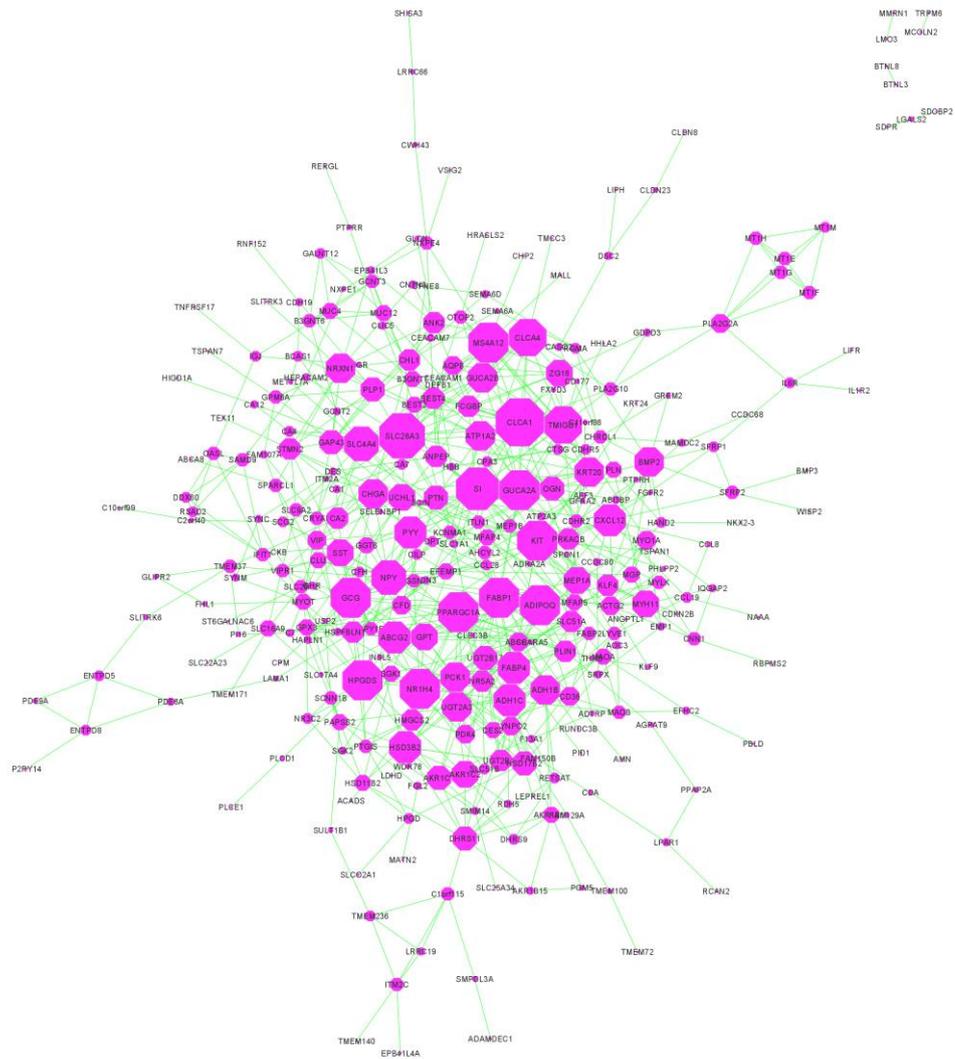

**Figure1: The protein-protein network in colon cancer. The pink nodes represent DEGs in colon cancer. The edges represent the interactions between them and size of pink nodes is proportional to Degree centrality of the proteins.**

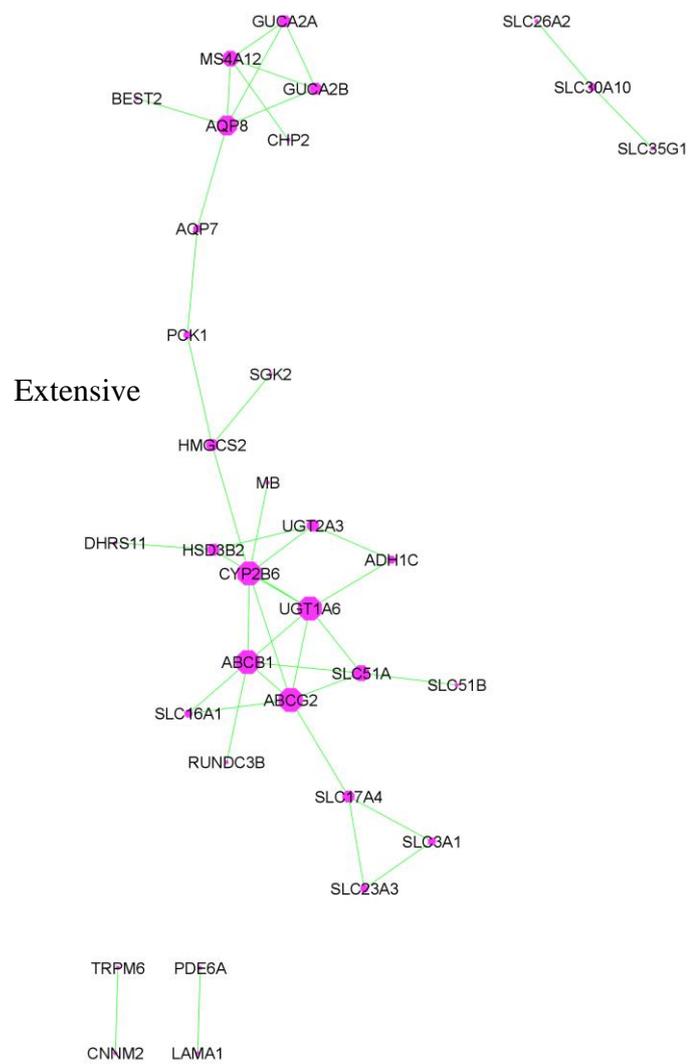

Extensive

Figure2: The protein-protein network in extensive ulcerative colitis. The pink nodes represent DEGs in extensive ulcerative colitis. The edges represent the interactions between them and size of pink nodes is proportional to Degree centrality of the proteins.

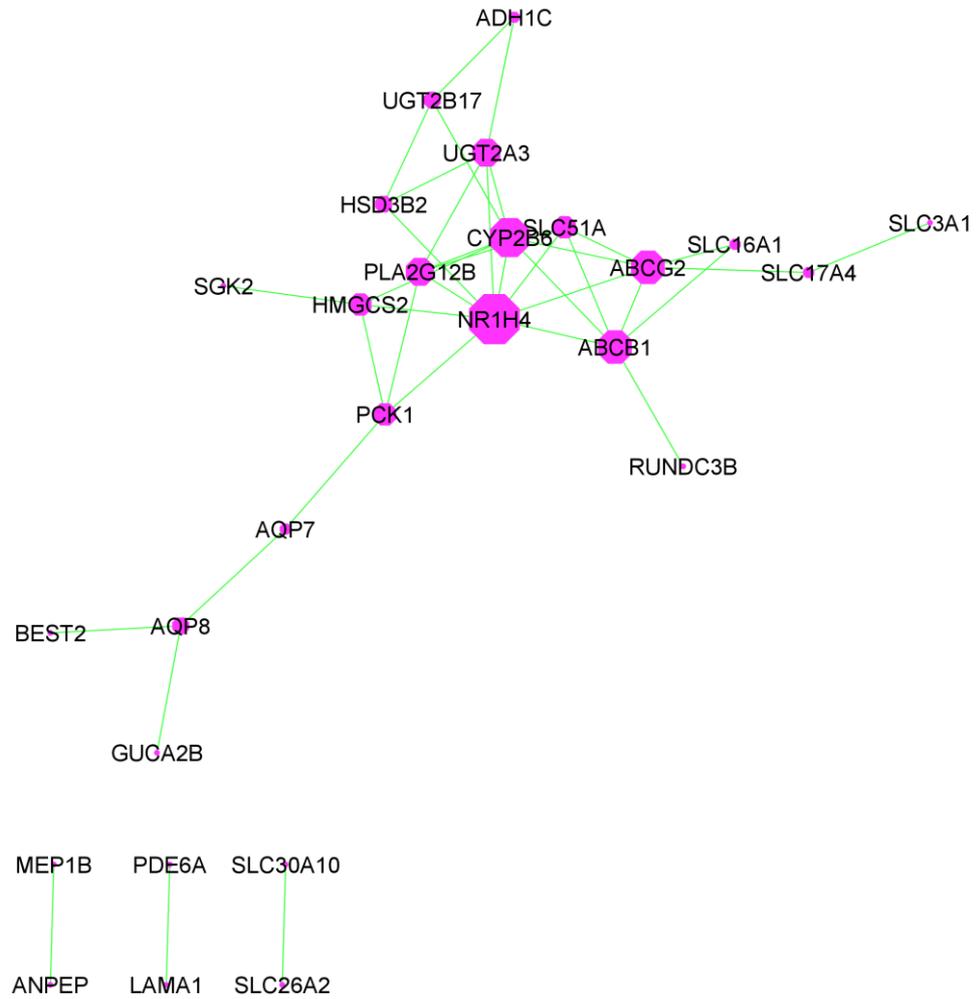

**Figure3: The protein-protein network in limited ulcerative colitis. The pink nodes represent DEGs in limited ulcerative colitis. The edges represent the interactions between them and size of pink nodes is proportional to Degree centrality of the proteins.**

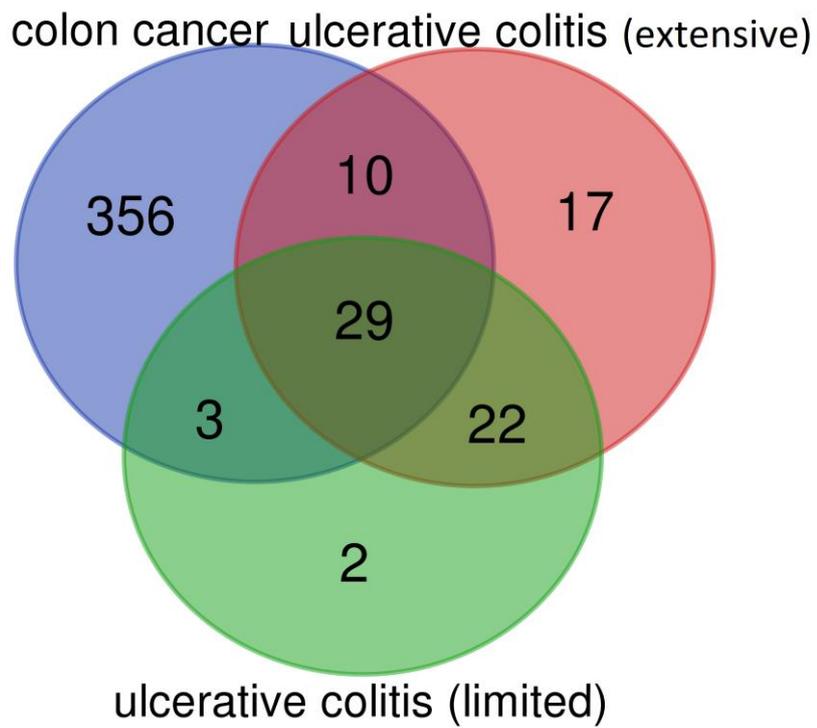

**Figure4: Identification of the shared and specific under-expressed DEGs in colon cancer, extensive ulcerative colitis and limited ulcerative colitis (based on inclusion criteria) via a Venn diagram. The blue circle indicates colon cancer, the green circle indicates limited ulcerative colitis, and the red circle indicates extensive ulcerative colitis.**

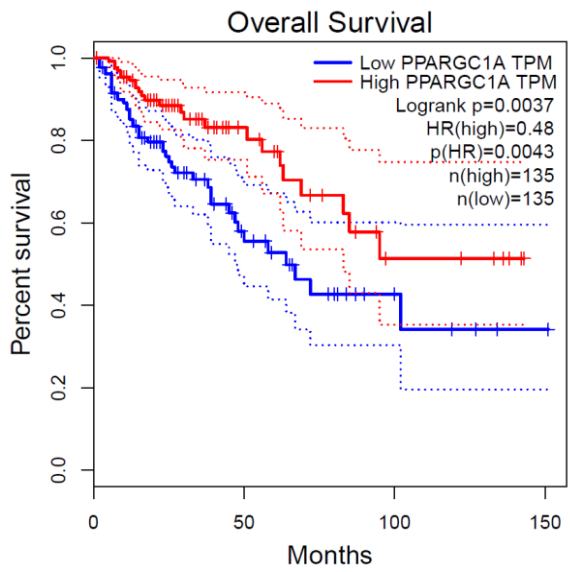
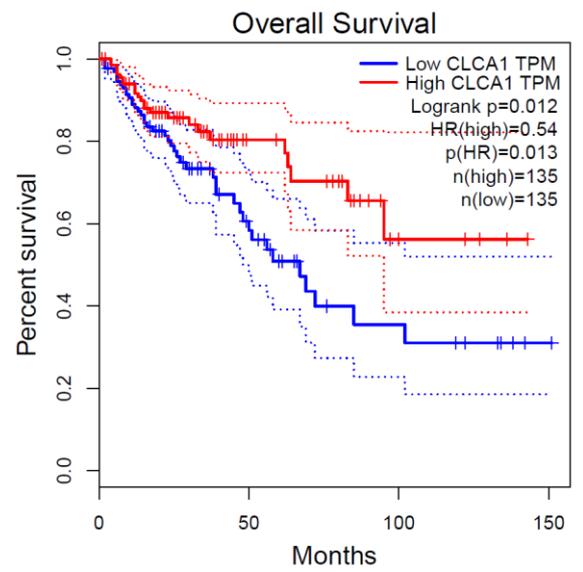
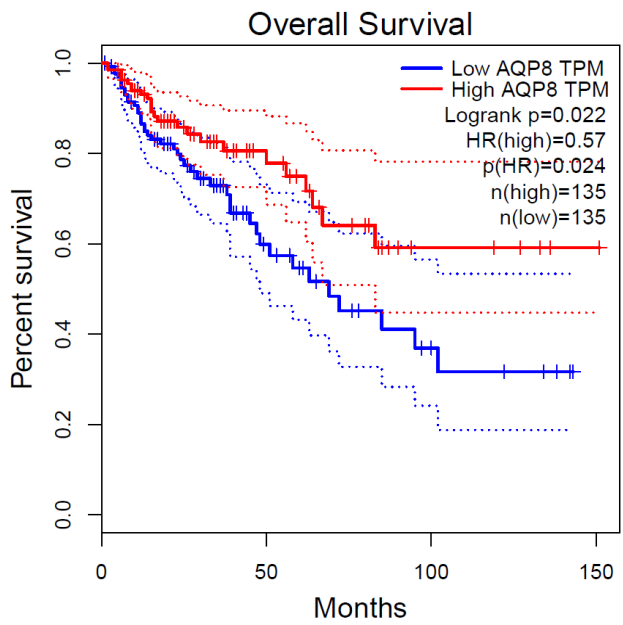

**Figure 5: Overall survival of key genes in colon cancer, extensive and limited ulcerative colitis was evaluated by Kaplan-Meier curve. CLCA1, PPARGC1A and AQP8 were with poor overall survival.**